# Bias-triggered conductivity relaxation (BCR): a unique tool to simultaneously investigate thermodynamics, kinetics and electrostatic effects of oxygen reactions in MIEC thin films

Alexander Stangl,[1,2,3*] Alexander Schmid,[4] Adeel Riaz,[3] Jürgen Fleig[4] and Arnaud Badel[5]

* alexander.stangl@tuwien.ac.at
[1] Atominstitut, TU Wien, 1020 Vienna, Austria
[2] Université Grenoble Alpes, CNRS, Grenoble INP, Institut Néel, 38000 Grenoble, France
[3] Université Grenoble Alpes, CNRS, Grenoble INP, LMGP, 38000 Grenoble, France
[4] Institute of Chemical Technologies and Analytics, TU Wien, 1060 Vienna, Austria
[5] Université Grenoble Alpes, CNRS, Grenoble INP, G2ELab – Institut Néel, 38000 Grenoble, France

## Abstract:

Mixed ionic electronic transfer (MIET) reactions, such as the oxygen reduction reaction (ORR) at oxide surfaces, are of paramount importance to manifold technologically highly relevant processes and fundamental understanding must be developed to improve performance and tailor highly efficient electrodes and catalysts. Understanding such complex multi-step reactions, requires the study of kinetic processes, underlying thermodynamic properties, *i.e.* ionic and electronic defect concentrations and electrostatic surface effects. However conventional techniques struggle to uncover the complete picture within the same sample/measurement. Here, we overcome this limitation by introducing bias-triggered conductivity relaxation (BCR) as a novel tool to investigate MIET reactions on oxides. It is based on alternating out-of-plane coulometric titration/polarization and in-plane electrical conductivity relaxation measurements, providing simultaneous electronic, ionic and extraordinarily rich surface kinetics information. This innovative combination of electrical and chemical driving forces synergizes information depth, with enhanced time resolution, versatility and speed, yet it lifts the weaknesses of the individual approaches, while remaining cost-effective and surprisingly simple. Furthermore, BCR allows to disentangle overpotential induced electrostatic modifications of the surface kinetics in a unique manner. We showcase the advantages of BCR in this work by studying the ORR in model $(La,Sr)FeO_{3-\delta}$ thin film electrodes and reporting on their thermodynamic and kinetic properties.

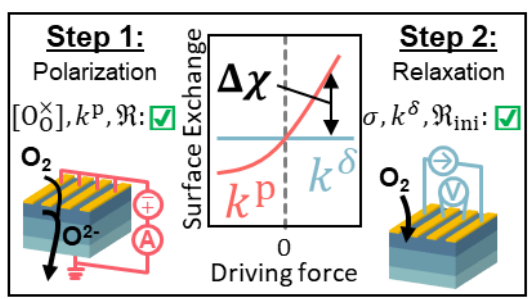

**Graphical abstract**

# 1. Introduction:

From fusion enabling superconductors[1] to renewable energy storage solutions based on highly efficient solid oxide fuel cells and batteries,[2,3] mixed ionic electronic conducting (MIEC) oxide materials are central to various emerging technologies to foster the energy transition and combat anthropogenic climate change.[4] In this context, oxygen transport reactions are twofold important. First, functional properties of oxides are highly influenced by their oxygen stoichiometry, making the precise control of the oxygen content key for superior materials performance,[5,6] while on the other hand, the operation principle of various energy related electrochemical processes is based on the oxygen reduction reaction (ORR) itself.[7] The accelerated development of tuned metal oxides relies on innovative approaches to materials synthesis and nano-engineering,[8] the improved understanding of ion kinetics and point defects,[9] and on a constantly expanding experimental toolbox.[10]

The ORR describes the incorporation of a gaseous oxygen molecule, $O_{2,gas}$, into two bulk crystal sites. For an oxygen vacancy-based defect chemistry it can be written as:

$$O_{2,gas} + 2v_O^{\bullet\bullet} \rightarrow 2O_O^{\times} + 4h^{\bullet} \qquad \text{Eq. 1}$$

with the oxygen vacancy, $v_O^{\bullet\bullet}$, the electronic hole, $h^{\bullet}$, and the filled bulk oxygen site, $O_O^{\times}$. In contrast to its apparent simplicity, the complexity of the ORR cannot be overstated, as the arrow in Eq. 1 covers a succession of elementary reaction steps, including adsorption, ionic and electronic charge transfer reactions across gas-solid electrochemical interfaces, dissociation, recombination with a crystal site and various diffusion processes in the presence of mixed electrostatic and chemical gradients.[11] These elementary mechanisms may occur several times in an *a priori* unknown order, while it is generally considered, that one of these fundamental reaction mechanisms is limiting the overall reaction rate, thus being the rate determining step (RDS). Identifying this rate limiting reaction of the ORR is an essential step towards materials-by-design with enhanced electrochemical performance, durability and possibly improved sustainability with reduced critical raw materials consumption. Countless efforts have been undertaken to shed light onto the RDS.[12–17] Given the apparent difficulties to follow these mixed ionic electronic transfer (MIET) reactions at the atomic and sub-atomic level,[18] macroscopic approaches, based on the analysis of pressure dependencies, have crystallized to be the most promising ones.[11] Oxygen transport processes are frequently studied using electrochemical impedance spectroscopy (EIS),[19] isotope tracer profiling using secondary ion mass spectroscopy (SIMS),[20–22] relaxation experiments, such as electrical conductivity relaxation,[23] or *in situ* XRD,[24] as well as dynamic equilibrium approaches, including electrochemical titration and current density-overpotential ($j - \eta$) measurements,[25] which provide complementary information on the inquired system. A general rate equation for the rate determining step was formulated in Ref. 26 and can be written in the limit of low coverage as:

$$\mathfrak{R} = \vec{\mathfrak{R}} - \overleftarrow{\mathfrak{R}}$$

$$\vec{\mathfrak{R}} \propto \underbrace{[pO_2]^{\frac{1}{n}}}_{\text{Adsorbates}} K_{eq} \underbrace{\prod_i [i]^{v_i}}_{\text{Defects}} \times \underbrace{e^{\frac{\vec{\beta} e \chi}{k_B T}}}_{\substack{\text{Electrostatic} \\ \text{term}}} \times \underbrace{e^{\frac{\vec{\beta}' e \Delta \chi(\eta)}{k_B T}}}_{\substack{\text{Overpotential} \\ \text{effects}}} \qquad \text{Eq. 2}$$

$$\overleftarrow{\mathfrak{R}} \propto \underbrace{\prod_j [j]^{v_j}}_{\text{Defects}} \times \underbrace{e^{\frac{\overleftarrow{\beta} e \chi}{k_B T}}}_{\substack{\text{Electrostatic} \\ \text{term}}} \times \underbrace{e^{\frac{\overleftarrow{\beta}' e \Delta \chi(\eta)}{k_B T}}}_{\substack{\text{Overpotential} \\ \text{effects}}}$$

The forward and backward rates, $\vec{\mathfrak{R}}$ and $\overleftarrow{\mathfrak{R}}$ constitute of contributions from adsorbates, ionic and electronic defect species, $i$ and $j$, with their respective reaction orders $v_i$ and $v_j$, as well as electrostatic effects, due to the presence of a surface potential step, $\chi$ and its variation under an applied overpotential, $\Delta\chi(\eta)$. The latter term subsumes electrostatic effects arising with the application of an overpotential, such as changes of the surface dipole layer, *e.g.* by a modified coverage of charged adsorbate species, which is also reflected in changes of the work function.[27–29] Defects can appear directly in the reaction rate if they participate in the RDS or indirectly via the equilibrium constant, $K_{eq}$. Note that defect concentrations and consequently $K_{eq}$ depend as well on the overpotential. Sophisticated measurement designs, as developed for example by Merkle[30], Schmid[31] and Guan[32] for chemical and electrical experiments, allow to separate and eliminate certain defect contributions in Eq. 2. Nevertheless, elucidating the ORR reaction pathway requires access to both, kinetic and thermodynamic materials parameters, including defect concentrations as well as reaction rates and exchange coefficients and their variations under current. Up to now, simultaneous determination of all these properties was not feasible, despite advanced approaches based on the integration of complementary techniques.[33–35] Thus, the mechanistic interpretation of kinetic processes is frequently based on different assumptions (*e.g.* Brouwer approximation), (bulk) literature data and/or several experiments under different conditions with samples subject to evolution, and may therefore not be sufficiently accurate for the specific investigated material (*e.g.* thin films).

Here, we aim to tackle this issue by combining, in an alternating operation mode, out-of-plane electrochemical titration and current-overpotential measurements with in-plane electrical conductivity relaxation (ECR) measurements. This approach is simple and easy to implement, yet allows to obtain all relevant information within a single experiment, including the ionic and electronic defect state of MIEC oxides, as well as kinetic parameters, such as the surface exchange coefficient and reaction rates. Furthermore, limitations of the individual techniques (such as flush time, accessible $pO_2$ range, inherent electrostatic modification of the surface) are lifted, while their information depths are synergistically combined. This provides enriched and novel insights and the unmatched opportunity to analyse bias-induced, electrostatic modifications of the reaction rates by comparing chemical and electrical experiments. The full potential and the substantial advantages of this new technique are demonstrated in the following by studying (La,Sr)FeO$_{3-\delta}$ thin films electrodes.

## 2. Methods:
### 2.1. Key concepts of known $j - \eta$ and ECR measurements

Time dependent current density-overpotential ($j - \eta$) measurements (which include coulometric titration) and electrical conductivity relaxation (ECR) are known standard approaches for the investigation of oxygen kinetics and defects in MIEC materials. Their basic measurement principles are summarized in Figure 1 and in the following.

Insight into reaction kinetics is typically obtained via an intentionally introduced (small) distortion of the equilibrium, *e.g.* in the case of current density-overpotential/coulometric titration measurements via the application of an electrical potential across an electrochemical cell. That is a pure ionic conducting electrolyte sandwiched between a MIEC material of interest as working electrode (WE) and a highly active counter electrode (CE), as illustrated in Figure 1(a). If WE surface reactions are slow compared to the oxygen transport properties of the remaining device components (including oxygen diffusion in the WE), a DC voltage applied across the cell translates (partially) into an overpotential, $\eta$, which modifies the oxygen chemical potential in the entire WE according to:

$$\mu_{O,WE} = \mu_{O,CE}(pO_{2,ref}) + 2e\eta \qquad \text{Eq. 3}$$

with $\mu_{O,CE}(pO_{2,ref})$ denoting the chemical potential of oxygen at the CE (being in equilibrium with the oxygen partial pressure of the surrounding gas phase $pO_{2,ref}$). Note that despite being an electrical experiment, there is a true chemical potential gradient involved. The oxygen stoichiometry of the MIEC is changed by pumping oxygen ions from/into the MIEC via the electrochemical cell, which resembles the building up of a cell voltage in a battery during charging. The time evolution of the $\mu_O$ across the gas-MIEC interface is drafted in Figure 1(c) (left panels). The four stages correspond to: (i) the MIEC is in equilibrium with the atmosphere prior to the application of a voltage ($t < 0$). (ii) A negative bias, $U_{DC} < 0$, results in a drop of $\mu_O$ over the RDS, which (iii) rapidly increases with time until (iv) the steady state is reached with $\Delta\mu_O = 2e\eta$. The new oxygen chemical potential inside the MIEC can be related to an effective oxygen partial pressure, $pO_{2,\text{eff}}$ (assuming ideal gas behaviour) using the Nernst equation:

$$pO_{2,\text{eff}} = pO_{2,\text{ref}} e^{\frac{4F\eta}{RT}} \qquad \text{Eq. 4}$$

with $F$ and $R$ being the Faraday and gas constants, respectively. Essentially as a second order effect of the applied voltage, the jump in $\mu_O$ drives a continuous net oxygen flux across the surface, corresponding to a leakage current. The saturation current density, $j_{\text{sat}}(\eta)$, in the steady state, is thus a measure of the net surface reaction rate, $\mathfrak{R}$:

$$\frac{j_{\text{sat}}(\eta)}{2e} = \mathfrak{R} = \vec{\mathfrak{R}} - \overleftarrow{\mathfrak{R}} \qquad \text{Eq. 5}$$

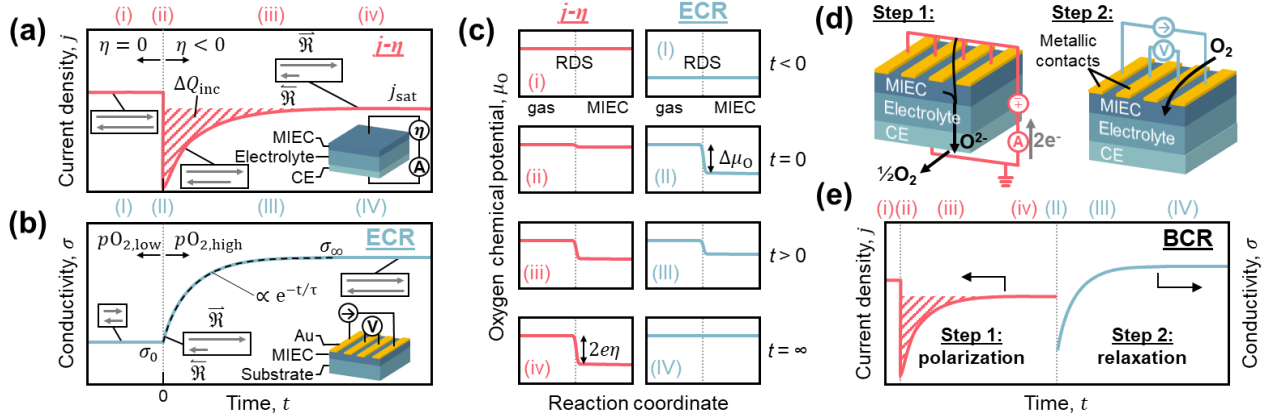

Figure 1: Overview of the measurement principle for (a) current density – overpotential ($j - \eta$) and (b) electrical conductivity relaxation (ECR) measurements with sketches of the sample configurations. The different stages in (a) and (b) are labelled with majuscule and minuscule Roman numerals, respectively: (i & I) initial equilibrium, (ii & II) perturbation of equilibrium via application of $\eta$ or jump in $pO_2$, (iii & III) relaxation towards the new equilibrium/steady state (iv & IV). Grey arrows in (a) and (b) illustrate forward and backward reaction rates. The MIEC undergoes a stoichiometric change, quantifiable via the marked red area. The oxygen chemical potential profiles across the surface of the MIEC for the different stages (i – iv and I – IV) are drafted in (c). (d) Special device architecture enables two different cell configurations. Fast switching between the electrode configurations is enabled via a digital relay. (e) Principle of novel bias-triggered conductivity relaxation (BCR) technique based on combination of polarization (step 1) and relaxation (step 2) measurements.

As this current consumes a fraction of the applied voltage, the attainable overpotential in the WE reduces to:

$$\eta = U_{DC} - j_{\text{sat}} A_{\text{film}} R_{\text{CE+EL}} \qquad \text{Eq. 6}$$

with the surface area of the sample, $A_{\text{film}}$, and the summed resistance, $R_{\text{CE+EL}}$, of all cell components but the working electrode. From the flown charge, i.e. from the time-dependence of the current, we can extract defect chemical thermodynamic information as for usual coulometric titration studies. Assuming a constant oxygen reaction rate at the surface and thus a constant leakage flux throughout the entire titration process, i.e. $j_{\text{leak}}(t) = j_{\text{sat}}$, we can quantify the relative change in oxygen off-stoichiometry, $\Delta\delta$, of the WE via the area under the $j(t)$ curve (marked red in Figure 1(a)):

$$\Delta\delta = \frac{V_{uc}}{2e \cdot d_{film}} \int (j(t) - j_{sat})dt \qquad \text{Eq. 7}$$

with the sample thickness, $d_{film}$, and the unit cell volume $V_{uc}$. Thus, we get information as for pure coulometric titration, *i.e.* with negligible leakage current. More realistically, the leakage current density is not constant but increases gradually with $\mu_{O,WE}$ (and thus with the cell voltage). Hence, $\Delta\delta$ as given in Eq. 7 acts more as lower bound, while its uncertainty increases with the magnitude of $j_{sat}$ (relative to $j(t)$) and the step size of $\eta$. However, for a rate-limiting surface step being much more resistive than all other kinetic processes in the cell (reaction at the CE, ion transport in the electrolyte, charge transfer into the MIEC, transport in the MIEC) the correction for $j_{sat}$ in Eq. 5 is small (a few %), as shown in the Supplementary Information Note 3, and meaningful non-stoichiometry data can be obtained despite a non-constant dc current flow.

A major drawback of polarization measurements is the inherent, very non-trivial coupling of the applied bias with charged surface defects, the electrostatic surface potential and the Fermi level of the bulk (and thus the work function). This intrinsically causes a modification of the reaction rate of the RDS and severely complicates the mechanistic interpretation of MIET reactions at gas-solid interfaces.

In conventional ECR experiments on the other hand, kinetic processes are triggered directly via a jump in the oxygen partial pressure surrounding the sample, while monitoring the conductivity, $\sigma$, of the material. These conductivity relaxation curves can be analysed using solutions of Fick's diffusion model to obtain the chemical surface exchange coefficient, $k^\delta$, and/or the chemical diffusion coefficient, $D^\delta$.[36] Accurate determination of transport parameters, however, requires that the gas exchange of the atmosphere is much faster than the kinetics of the sample, to avoid interference of these two processes.[37] Ideally, the $pO_2$ vs. time curve follows a square wave (*i.e.* instantaneous partial pressure jump), which is experimentally difficult to achieve and limits its applicability to temperatures, where kinetics are sufficiently slow.

An exemplary oxidation step is shown in Figure 1(b): (I) in the initial state at low $pO_2$, forward and backward reaction rates are in equilibrium (grey arrows). (II) With the rise of the $pO_2$, $\overrightarrow{\mathfrak{R}}$ jumps to the new value (due to its direct dependence on $pO_2$), while (III) $\overleftarrow{\mathfrak{R}}$ only increases gradually. This leads to a net flux of oxygen, $J(t)$ (equal to the net reaction rate, $\mathfrak{R}(t)$), into the MIEC, which vanishes once the new equilibrium state is reached (IV). Within a first order, linear kinetic regime, the exchange flux through the surface is proportional to the oxygen concentration difference between the first surface layer, $c(t)$ and the value corresponding to equilibrium with the gas phase ($c_\infty$):

$$J(t) = k^\delta(c_\infty - c(t)). \qquad \text{Eq. 8}$$

with the proportionality constant being $k^\delta$. For a surface limited thin film within the plane-sheet approximation and a linear relation between $\sigma$ and $c$, the normalised conductivity transient can be modelled using a single exponential function:

$$\sigma_{norm} = \frac{\sigma(t) - \sigma_0}{\sigma_\infty - \sigma_0} = 1 - e^{-\frac{t}{\tau}} \qquad \text{Eq. 9}$$

with the time constant, $\tau = d_{film}/k^\delta$. The evolution of the $\mu_O$ profile is drawn in Figure 1(c) (right panels). Prior to a forward jump, the initial, low $\mu_O$ is constant across the surface. In the ideal case, at $t = 0$ a step in $\mu_O$ builds up instantaneously with the step size $\Delta\mu_O = RT/2 \log(pO_{2,low}/pO_{2,high})$. With increasing time, $\Delta\mu_O(t)$ decreases, as the new equilibrium is approached. For a surface limited reaction, $\mu_O$ is homogeneous and in dynamic quasi-equilibrium inside the MIEC at all times.

Both methods lead to kinetic information on the oxygen exchange process at the WE surface, namely $j_{\text{sat}}$ in $j-\eta$ studies and $k^\delta$ in ECR measurements. Interestingly, the measurement times required to get these complementary surface kinetic data are generally rather different. This is due to the fact that in time-dependent $j-\eta$ studies the chemical potential changes in the MIEC takes place via an "out-sourcing" of the ORR to the much faster CE. Hence, all the (smaller) resistances except the rate-limiting surface step are decisive, while for ECR the rate determining surface step limits the equilibration, which thus becomes slower. In other words, the chemical capacitance of the WE gets charged via the (generally small) $R_{\text{CE+EL}}$ in $j-\eta$ measurements and discharged via the (larger) $R_{\text{WE}}$ in ECR experiments. Moreover, the analysis of the time-dependent current in $j-\eta$ measurements allows to quantify the defect chemical state of the MIEC, which is otherwise challenging to obtain, while the ECR method provides insight on the material's electronic state.

### 2.2. Novel bias-triggered conductivity relaxation (BCR)

Here, we propose to merge current density-overpotential measurements (including coulometric titration) and electrical conductivity relaxation into a single methodology, bias-triggered conductivity relaxation (BCR), to synergize their individual merits and overcome their inherent weaknesses. This ingenious combination ultimately provides highly efficient access to all relevant parameters for the in-depth characterisation of MIET reactions in oxides and opens up many novel considerations. In comparison to reported, joint titration-conductivity approaches, relaxation and titration measurements are directly performed on the same specimen,[38] the kinetic dimension is analysed rather than suppressed,[39–42] and electrochemical experiments are matched against chemical ones,[35] which significantly enriches our results and simplifies setup and sample preparation.

For BCR measurements we deploy a specific electrochemical cell architecture with four metallic top electrodes, as shown in Figure 1(d) and to our knowledge originally introduced by Yugami *et al*.[42] In a first step, the metallic top electrodes are short-circuited and the sample is biased across the electrolyte to establish an effective $pO_2$ inside the MIEC, equivalent to $j-\eta$ measurements, see Figure 1(e). This step is followed by a rapid, digital relay-controlled switch of the electrode configuration to cut the out-of-plane connection and contact the top electrodes individually for in-plane, four-point resistivity measurements. Upon removal of the overpotential, the effective $pO_2$ can no longer be retained within the MIEC and the system relaxes back to the thermodynamic equilibrium state defined by the $pO_{2,\text{ref}}$. Cutting the external out-of-plane connection ensures that any ionic flow through the electrolyte is stalled and relaxation takes place exclusively via the native MIEC top surface.

Compared to standard ECR measurements, the step in the $pO_2$ of the gas phase is replaced with a virtually instantaneous electrochemical step in $pO_{2,\text{eff}}$, whereas the initial and final state are defined by $\eta$ (and the corresponding $pO_{2,\text{eff}}$ via the Nernst equation, Eq. 4) and $pO_{2,\text{ref}}$, respectively. The equivalence of this can be easily understood by looking at the schematics in Figure 1(c), where the ECR phase (II) directly results from the polarization phase (iv). As a jump in $pO_2$ is omitted, this approach is free of any flush time limitation and enables to operate with reduced gas flow rates or even under static conditions, decreasing surface cooling effects and providing strongly enriched flexibility for the study of different atmospheric compositions and $pO_{2,\text{eff}}$ step sizes. Note that, although the MIEC WE is effectively reduced upon application of a cathodic bias, oxygen is incorporated at the WE surface (oxidation reaction). In the subsequent relaxation step the reduced MIEC is re-oxidized. Thus, the net oxygen flux across the WE surface is in the same direction during step 1 and 2, as illustrated in Figure 1(d) (see black arrows). We therefore use the terms oxidation and reduction to refer to the WE surface processes during and following cathodic ($\eta < 0$) and anodic ($\eta > 0$) polarizations, respectively.

Altogether, BCR delivers several important kinetic parameters (chemical, electrical and electrochemical surface exchange coefficients, as well as initial and net reaction rates) of one and the same surface (including their activation energies and reaction orders) and a complete picture of the ionic and electronic defect state of the MIEC ($\Delta\delta, \sigma$) within a single experiment. At the same time, inaccuracies due to finite flush times or not well-defined starting points of the conductivity relaxation process are avoided and the influence of voltage-induced modulations of the surface potential, $\Delta\chi(\eta)$, on the reaction kinetics can be decoded in a unique manner by comparing kinetic parameters of electrochemical and chemical origin from complementary polarization and relaxation measurements. Furthermore, thorough BCR data analysis allows to separate kinetic contributions of the remaining cell components (CE + electrolyte) and quantify the voltage-overpotential conversion efficiency, covering aspects of EIS. All this is demonstrated in the following section.

Moreover, BCR captivates by its very simple setup, which consists only of a temperature system (with five electronic contacts), a standard source meter and a digital relay. This will benefit its fast dissemination and implementation as new, fast and cost-effective standard tool, which can be readily coupled to other *in situ* techniques such as optical absorption, NAP-XPS, X-ray diffraction, spectroscopic ellipsometry and Raman spectroscopy to provide additional insights on specific reaction intermediates, the structural and surface chemical evolution, *etc*. Brief amendatory notes on sample and setup requirements for BCR measurements can be found in the Supplementary Information Note 1.

## 3. Results and Discussion

We have studied dense, 120 nm thick sub-stoichiometric (La,Sr)FeO$_{3-\delta}$ thin films deposited by pulsed laser deposition (PLD) on 5×5 mm ionic conducting La$_{0.95}$Sr$_{0.05}$Ga$_{0.95}$Mg$_{0.05}$O$_{3-\delta}$ (LSGM) single crystal substrates (100). SEM images of the top surface and XRD pattern of a fully reduced and fully oxidised sample can be found in SI-Figure 2, revealing small grain sizes, with an average equivalent diameter of about 80 nm, and a (h00) pseudo-cubic structure.[43] Four 100 nm thick Au top electrodes have been fabricated using photolithography and metal e-beam evaporation in cleanroom facilities. A porous, paint brushed Ag layer served as counter electrode. We verified that WE surface reactions are limiting the overall oxygen pathway and thus $\eta \approx U_{\mathrm{appl}}$ using electrochemical impedance spectroscopy (EIS) and by comparing conductivity and incorporated charge for different effective and reference oxygen partial pressures, see the Supplementary Information Note 2 and SI-Figure 3(a, c & d). Additionally, we performed gas sensor type measurements, where WE and CE are electrically short-circuited using an ammeter and the out-of-plane current is measured upon performing a switch in the gas atmosphere, as shown in SI-Figure 3(b) for a jump towards lower $pO_2$. The negative sign of the current implies that oxygen ions are transported from the WE to the CE through the electrolyte. Thus, the oxygen stoichiometry of the MIEC equilibrates to the lower $pO_2$ predominantly through the CE rather than through its native surface, as shown by the analysis of the transported charge in the Supplementary Information. The fact that the oxygen evolution reaction is "out-sourced" to the CE surface confirms that surface reactions of the WE are limiting the overall equilibration process, which is a precondition for the following BCR measurements, performed on LSF thin films.

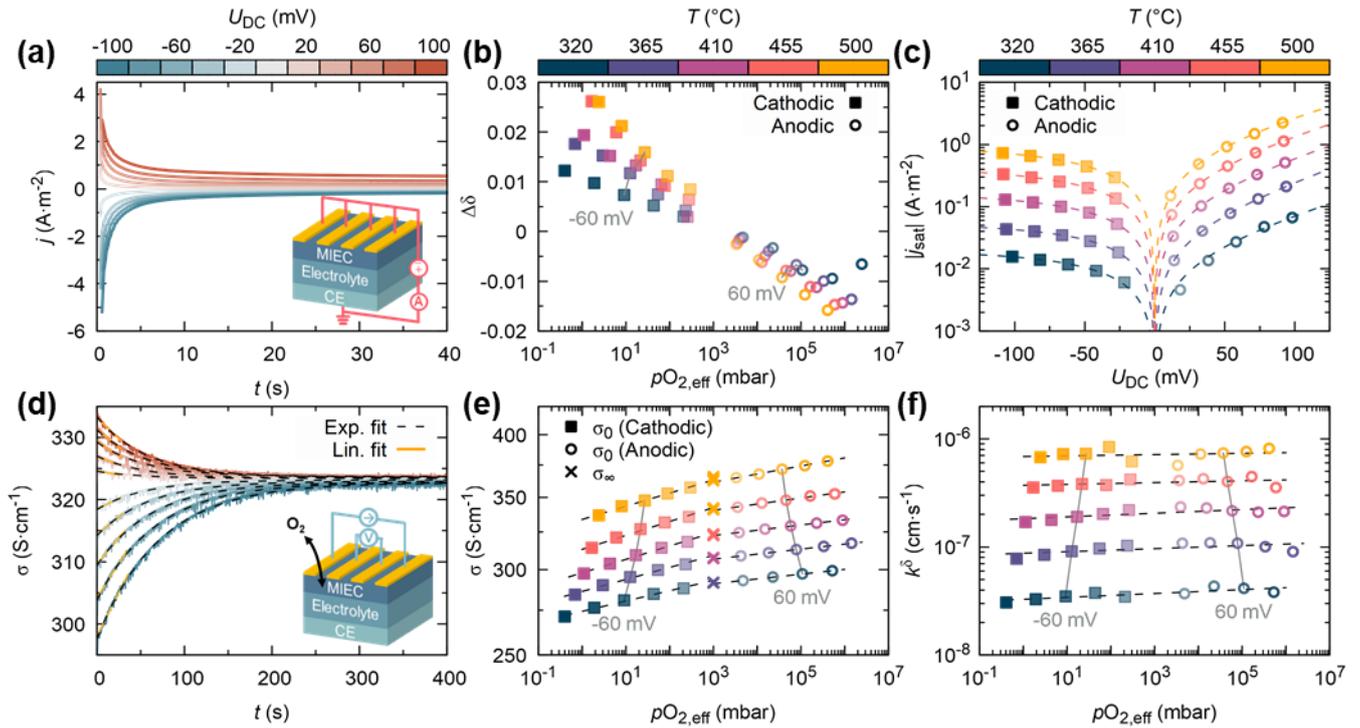

Figure 2: Bias-triggered conductivity relaxation (BCR) measurements: (a) polarization curves (step 1) and (d) subsequent conductivity relaxation processes (step 2, colouring according to step 1) at 410 °C in 1 bar of flowing oxygen for various positive and negative biases. (b) Change in oxygen stoichiometry as function of effective oxygen pressure and (c) saturation current density as function of applied voltage, obtained from polarization curves at different temperatures. Log-log plots of the (e) initial and final conductivity, $\sigma_0$ and $\sigma_\infty$, respectively and (f) the chemical surface exchange coefficient as function of the initial effective $pO_2$ inside the MIEC for various temperatures ($pO_{2,\text{final}} = 1000$ mbar). Dashed lines are guides to the eye. Values in (e & f) were extracted from relaxation data shown in (d) using exponential fitting curves (black dashed lines). The magnitude of the (preceding) bias is indicated via the symbol's opacity in (b, c, e & f).

### 3.1. "1+1=2": added up information depth using BCR

The integration of different techniques into a single BCR experiment allows to simultaneously determine all relevant materials descriptors for MIET reactions. The rich and abundant information density provided by BCR measurements is summarized in Figure 2, including thermodynamic and kinetic parameters obtained from polarization (top row) and subsequent relaxation (bottom row) measurements.

**In a first step**, the application of a bias, $U_{\text{DC}}$, triggers a current of same polarity, which rapidly decays, as shown in Figure 2(a) for biases between -100 and +100 mV at 410 °C. Additional raw data for other temperatures can be found in SI-Figure 4. Commonly, two figures of interest are extracted from such plots, namely the charge incorporated into the WE, corresponding to the change in oxygen stoichiometry, $\Delta\delta$, and the saturation current density, $j_{\text{sat}}$, which is a measure of the net reaction rate. Additionally, we identified that the onset current density, $j(t = 0)$ contains entangled information about the kinetic activity of the remaining components of the electrochemical cell, including the electrolyte and the CE and as such, is a function of the voltage step. This relation can be used to analyse the process of building-up the overpotential inside the WE and quantify the voltage-overpotential conversion efficiency (here 95 %), as well as to estimate the error in the incorporated charge, due to a non-constant leakage current, see Supplementary Information Note 3 and the corresponding SI-Figure 5. A more detailed analysis will be the subject of a forthcoming paper.

The relative change in oxygen stoichiometry, $\Delta\delta = \delta(\eta \neq 0) - \delta(\eta = 0)$, obtained via Eq. 7, is plotted as function of the effective $pO_{2,\text{eff}}$ (Eq. 4), in Figure 2(b). The oxygen off-stoichiometry varies within a narrow window of -0.02 to 0.03, clearly showing that the overpotential is compensated ionically. In accordance with bulk behaviour,[44] the oxygen content increases with $pO_{2,\text{eff}}$ (*i.e.* $\Delta\delta$ decreases) and the slope steepens with increasing temperature.

The current density-voltage characteristics are plotted in Figure 2(c). An analytical expression for such curves was derived previously for certain limiting situations,[45] which is formally equivalent to an exponential Butler-Volmer type equation:

$$j = j_0 \left[ e^{\frac{\alpha_a F(\eta - U_0)}{RT}} - e^{\frac{-\alpha_c F(\eta - U_0)}{RT}} \right], \quad \text{Eq. 10}$$

with the exchange current density, $j_0$, and a voltage offset, $U_0$, which here likely corresponds to a thermovoltage due to asymmetric heating in the deployed temperature cell ($\lesssim 10\,\text{mV}$). The coefficients $\alpha_a$ and $\alpha_c$ for the cathodic and anodic branches are functions of the reaction orders of the defects participating in the forward and backward reactions of the ORR, respectively.[45] Good modelling can be achieved for all temperatures using least square fitting, as shown by dashed lines. The corresponding fit parameters are reported in SI-Figure 6(a & b). The exchange current density, $j_0$, rapidly increasing with temperature, is directly related to the electrical surface exchange coefficient, $k^q = j_0(2ec_O)^{-1}$, with the total oxygen ion concentration, $c_O$,[46] see SI-Figure 6(c). As shown in SI-Figure 6(a), $\alpha_a \approx 1$ and $\alpha_c \approx 0$ remain approximately constant with temperature, suggesting the predominance of a single RDS over the analysed temperature range. In contrast to simple cases in aqueous systems, however, $\alpha_c \approx 0$ does not imply a diffusion limited process but can be the result of two scenarios: either defects participating prior to or during the RDS of the ORR are independent of the overpotential or more likely, their individual contributions counterbalance each other.

**In a second step**, the applied overpotential is cut and the effective $pO_2$ state of the WE re-equilibrates with the surrounding atmosphere exclusively through its own surface. This relaxation can be followed electrically by rapid digital switching from the out-of-plane polarization to the in-plane resistivity electrode configuration, as sketched in the inset of Figure 2(d). Conductivity relaxation curves from various preceding anodic/cathodic biases at 410 °C are presented in Figure 2(d), for other temperatures see SI-Figure 4. The equivalence of BCR and true ECR relaxation transients, based on changes in effective $pO_2$ vs. atmospheric $pO_2$, is verified in SI-Figure 7.

As predicted above, the relaxation through the native WE surface is one to two orders of magnitude slower than the outsourced oxygen *charging* of the WE via the CE during the preceding polarization step. Transitions from higher to lower oxygen stoichiometries are accompanied by a decrease in conductivity (reduction, red curves) and vice-versa (oxidation, blue curves), as expected for a predominantly *p*-type conductor. All relaxation curves can be well modelled using the exponential Eq. 9 (dashed black lines in Figure 2(d)), providing access to the saturation time, $\tau$, as well as the initial and final conductivity values, $\sigma_0$ and $\sigma_\infty$. It is noteworthy that good fitting with a single saturation time is expected only for purely surface limited processes,[37,47] which in combination with shorter polarization than relaxation times, is an additional confirmation that a WE surface reaction limits the cell.

Saturation of all curves at the same $\sigma_\infty$ value (± 0.5 S·cm⁻¹) is a good indicator that there are no irreversible changes affecting the materials conductivity during the BCR measurements. Given the very fast electronic switching ($\ll 1\,\text{s}$) between out-of-plane polarization and in-plane electrical measurements, $\sigma_0$ approximately equals the equilibrium conductivity for the effective oxygen pressure inside the MIEC during the preceding polarization step. This is confirmed as well by the overlap of conductivity data at 5 and 1000 mbar in SI-Figure 3(c). The $pO_{2,\text{eff}}$ and temperature dependence of the electrical conductivity is shown in Figure 2(e) and SI-Figure 8(b), respectively. The measured $\sigma$ values compare well with literature for highly oxidised LSF.[40,48–50] The very low reaction order, $m(\sigma) = \frac{\partial \sigma}{\partial pO_2}$, of around 0.01 for the cathodic branch, which levels off even further to 0.005 for anodic polarisations (see SI-Figure 9(a)), is an indicator that the structure is approaching the stoichiometric endpoint with $\delta = 0$ at 1 bar, where the electron hole concentration is constant and given by the cationic substitution concentration. Going to more reducing biases (thus lower $pO_{2,\text{eff}}$) and/or lower atmospheric $pO_2$, the reaction order increases to about 0.15 (see SI-Figure 3(c)), becoming closer to reported LSF thin film data.[51] For bulk LSF under oxidising conditions, the pressure dependency is typically in the range of 0.2 to 0.25 above 600 °C.[49,52,53]

The chemical surface exchange coefficient is obtained via the time constant, $k^\delta = d_{\text{film}} \tau^{-1}$. Its thermally activated nature is readily visible in Figure 2(f), with a strong increase in magnitude going from 320 °C to 500 °C. In contrast to $j_{\text{sat}}$, $k^\delta$ is barely affected by the preceding bias with no significant asymmetry between anodic and cathodic branches, *i.e.* $k^\delta$ is independent of the $pO_2$ step size and its direction (within the studied range) and thus the initial defect state of the bulk, including oxygen vacancy and electron hole concentrations. This means that the duration of the relaxation process is independent of the total amount of oxygen to be incorporated, which confirms the validity of a first order kinetic regime (Eq. 8), *i.e.* the reaction rate is proportional to the change in oxygen concentration, $\Delta c = c_\infty - c(t)$. Reducing $pO_{2,\text{ref}}$ on the other hand, leads to a significantly lower $k^\delta$, as shown in SI-Figure 10(e), with a $pO_2$ reaction order, $k^\delta \propto pO_2^m$, of about $m \approx 0.3$ (see SI-Figure 10(f)). This highlights the strong influence of the $pO_{2,\text{ref}}$ dependent surface state and adsorbate coverage on $k^\delta$. Similar results have been discussed previously in literature, including matching $k^\delta$ values for oxidation and reduction steps in La$_{0.6}$Sr$_{0.4}$FeO$_{3-\delta}$ thin films,[51] as well as $k^\delta$ being independent of the step size $\Delta pO_2$ (while maintaining the final $pO_2$ constant) in La$_{0.5}$Sr$_{0.5}$CoO$_{3-\delta}$ bulk samples.[54,55]

By comparing $k^\delta$ and $k^q$, one can easily obtain the thermodynamic factor, which amounts to about 100 and therefore matches well the typical value for oxides with large chemical capacitance.[46] As an additional kinetic parameter, the initial rate of change, $\mathfrak{R}_{\text{ini}}$, can be obtained from the slope of the relaxation transients close to $t = 0$ (*cf.* linear fits with the slope $a$, given by solid orange lines in Figure 2(d)).[30] Alternatively, $\mathfrak{R}_{\text{ini}}$ can be estimated from the first term of the series expansion of the exponential fitting curve:

$$\mathfrak{R}_{\text{ini}} \propto a \approx \Delta \sigma \tau^{-1}, \qquad \text{Eq. 11}$$

with $\Delta \sigma = \sigma_\infty - \sigma_0$, as confirmed in SI-Figure 8(a). Before discussing $\mathfrak{R}_{\text{ini}}$ further in the next section, we analyse the temperature evolution of derived kinetic parameters.

Arrhenius plots for $k^\delta$ and $|j_{\text{sat}}|$ are presented in Figure 3(a & b), respectively. As noted above, $k^\delta$ values for oxidation (blue) and reduction steps (red) are basically ident and fall onto the same line, with matching activation energies of around 0.65±0.05 eV, as shown in in Figure 3(c). This value is within the wide range of reported activation energies for LSF, ranging from approximately 0.5 eV to 1.5 eV for different LSF compositions and sample types.[38,56–58] For $|j_{\text{sat}}|$ we find distinct activation energies for cathodic (blue) and anodic (red) polarizations, see Figure 3(d), both slightly higher than the activation energy for $k^\delta$ and closer to the activation energy of the electrical surface exchange coefficient, $k^q$. The pre-exponential factors from the corresponding Arrhenius laws for $|j_{\text{sat}}|$ and $k^\delta$ are given in SI-Figure 6(d), which in contrast to $E_A$ present a clear dependence on polarity and magnitude of the bias.

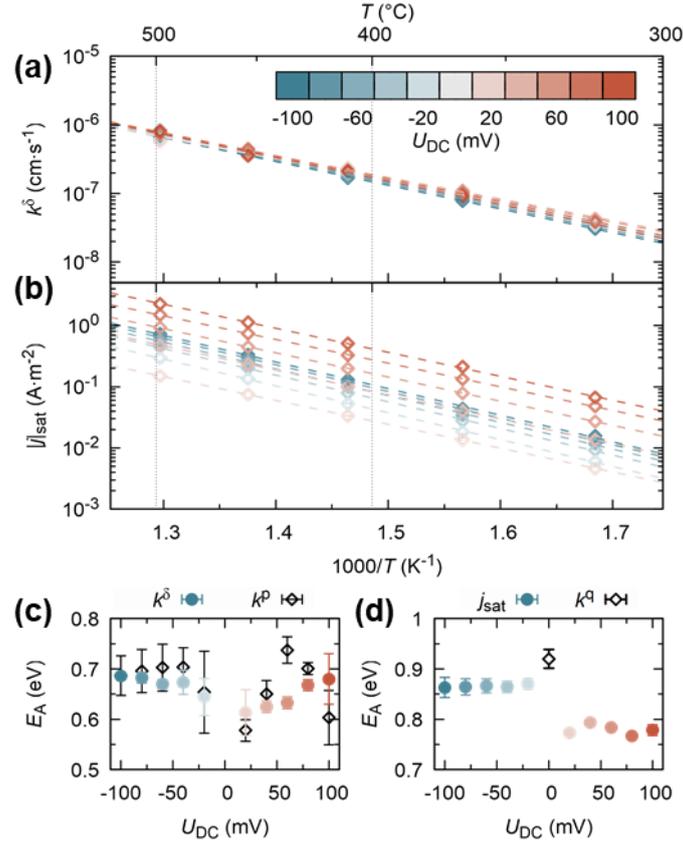

Figure 3: Arrhenius plots for (a) $k^\delta$ and (b) $|j_{sat}|$ for different polarizations at 1 bar of $O_2$. The corresponding activation energies are shown in (c) and (d), respectively. Panel (c) also displays the activation energy for the electrochemical surface exchange coefficient, $k^p$ (open diamonds), introduced and discussed in section 3.2.

### 3.2. "1+1=3": synergetic aspects of BCR

In the previous section we demonstrated the ability of BCR to inquire all relevant materials properties within a single measurement. However, the simultaneous determination of these quantities enables unique correlations and new insight and thereby exceeds the information depth of the individual approaches. These synergetic aspects of BCR are highlighted in the following.

For instance, by combining titration with conductivity measurements, BCR readily provides the dependence of electronic properties on relative changes in oxygen deficiency, not accessible via standard $\sigma$ versus $pO_2$ experiments (Brouwer analysis). Knowledge of such experimental $\sigma - \Delta\delta$ relationships, however, is essential for understanding the physical origin of changes in electronic conductivity as well as for the correct mathematical modelling of ECR transients (*e.g.* lack of this knowledge frequently limits ECR measurements to small $pO_2$ steps, where a linear $\sigma - \Delta\delta$ relation can be assumed[47]). At high temperatures, we find a linear $\sigma - \Delta\delta$ dependence, as shown in Figure 4(a). This suggests that variations in $\sigma$ are dominated by changes in the charge carrier density via the oxygen reduction reaction and that the hole mobility is constant and proportional to the slope. The emerging, weak deviation from linearity at lower temperatures results likely from changes in the charge carrier mobility, as reported previously for LSF.[59,60]

In addition to the standard kinetic analysis performed above, BCR enables the direct comparison of kinetic processes triggered by either purely chemical or electrical driving forces. This allows to unravel the intrinsic effects of an overpotential on the reaction kinetics under current. To the best of our knowledge, a similar approach has not yet been reported in literature.

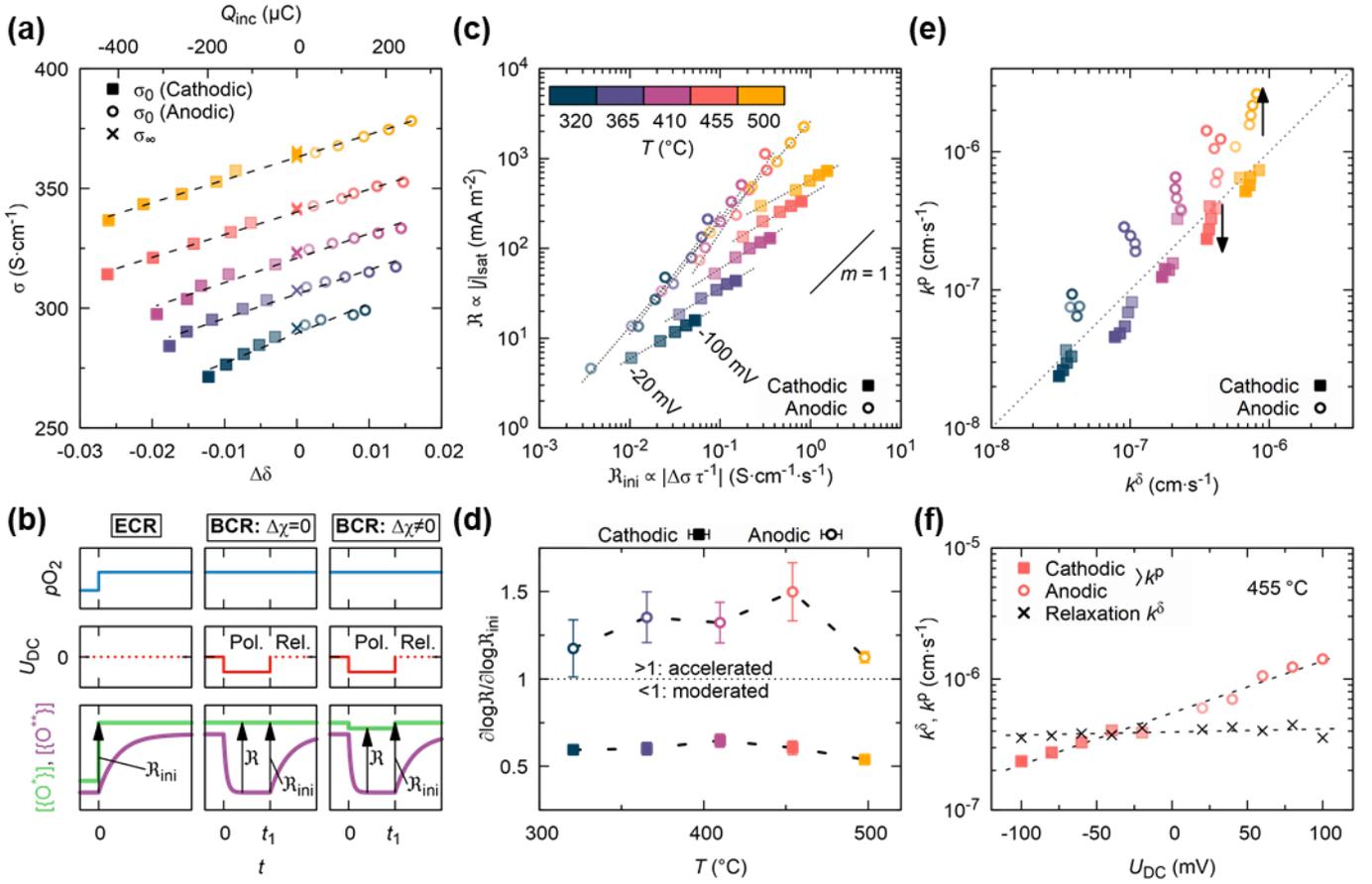

Figure 4: (a) Electronic conductivity of LSF as function of changes in oxygen off-stoichiometry. Dashed lines are guide to the eye. (b) Schematic time profiles of $pO_2$, $U_{DC}$ and defect concentrations of the oxygen intermediate states before and after the RDS, $\{O^*\}$ and $\{O^{**}\}$, respectively, for ECR and $j-\eta$ measurements. If $\{O^*\}$ carries electronic charge, its concentration is modified upon application of a bias and $\Delta\chi \neq 0$. A dotted line for the voltage indicates open circuit across the device. (c) Correlation between independent measures of the (initial) reaction rate obtained by $j-\eta$ (y-axis) and ECR (x-axis) measurements. Dotted lines are linear fits for cathodic and anodic branches at each temperature, with their slopes being shown in (d). (e) Direct numerical comparison of the electrochemical and chemical surface exchange coefficients, $k^p$ and $k^\delta$, respectively, obtained from $j-\eta$/titration measurements and the subsequent relaxation processes using the BCR methodology. (f) Bias dependence of $k^p$ and $k^\delta$ at 455 °C. The opacity of the symbols indicates the magnitude of the bias voltage in (a, c, e & f).

We start by looking at the reaction rates $\mathfrak{R}$ and $\mathfrak{R}_{ini}$ obtained from polarization and subsequent relaxation curves, respectively. By design of the experiment, the step in $\mu_O$ between the atmosphere and the MIEC bulk is identical for both reaction rates (*cf.* Figure 1). Consequently, any difference between $\mathfrak{R}$ and $\mathfrak{R}_{ini}$ must be directly linked to the applied voltage and the presence of electrical currents during the first step, due to bias-induced energy shifts of the work function, variations of the surface coverage of charged adsorbate species (*e.g.* $[O_{ad}^{2-}]$) and resulting changes of the electrostatic dipole layer of the work function, variations of the surface coverage of charged adsorbate species (*e.g.* $[O_{ad}^{2-}]$) and resulting changes of the electrostatic dipole layer. These effects ultimately alter the reaction intermediate concentrations, as well as the surface electrostatic potential step and thus the reaction rate.

Let us clear this point by re-analysing a traditional ECR step, as well as BCR measurements with and without modifications of the electrostatic surface potential. The corresponding $pO_2$ and $U_{DC}$ profiles (representing the driving forces), as well as defect concentrations of the reactant and product oxygen intermediate species of the rate determining step, $\{O^*\}$ and $\{O^{**}\}$, respectively, are illustrated in Figure 4(b). For an ideal ECR experiment, $[\{O^*\}]$ jumps instantaneously to the new equilibrium concentration, while at $t = 0$, $[\{O^{**}\}]$ is still in the pre-equilibrium state and starts transitioning to the new equilibrium with $\Re_{\text{ini}}$. For a BCR experiment with $\Delta\chi(\eta) = 0$, the applied overpotential only modifies $[\{O^{**}\}]$, resulting in a net reaction rate $\Re$. The steady state during polarization and the starting point ($t = 0$) of the subsequent relaxation have the same oxygen intermediate concentrations $[\{O^*\}]$ and $[\{O^{**}\}]$ and thus $\Re$ and $\Re_{\text{ini}}$ are equal, as indicated by arrows in the bottom centre schematic of Figure 4(b).

For sufficiently large deviations from equilibrium, *i.e.* large effective $pO_2$ steps, the net reaction rate can be approximated by the forward reaction rate, $\Re \approx \overrightarrow{\Re}$.[30] In such a case, $\Delta\chi(\eta) = 0$ is only expected, if no net electric charge is involved prior or during the RDS (that is $\{O^*\}$ is neutrally charged), and changes of the work function are neglectable. However, if $\{O^*\}$ is a charged state, *e.g.* adsorbed superoxide ($O_{2,\text{ad}}^-$) is involved as reactant in the RDS, an electric field and resulting currents can act on $\{O^*\}$, capacitively modifying its concentration and impede or enhance any additional charge transfer across the surface dipole layer during the RDS in a non-trivial manner. This will affect the reaction rate, $\Re$, as schematically drawn in Figure 4(b) (bottom right). Based on the RDS assumption, changes in $\{O^*\}$ and surface electrostatic relaxation processes take effect quasi instantaneously after cutting the applied voltage and $\{O^*\}$ relaxes immediately to the unbiased equilibrium with $\Delta\chi(\eta = 0) \equiv 0$. Thus, opposite to $\Re$ in the steady state, the initial reaction rate, $\Re_{\text{ini}}$, of the relaxation transient remains unaffected by the preceding bias and is equivalent for BCR and ECR relaxation steps.

The experimentally obtained reaction rates $\Re$ and $\Re_{\text{ini}}$ for LSF are compared in Figure 4(c). We find indeed a correlation between these quantities, with distinct power law dependencies $\Re \propto \Re_{\text{ini}}^m$ for the cathodic (oxidation) and anodic (reduction) branches, with $m < 1$ for cathodic and $m > 1$ for anodic biases at all temperatures, see Figure 4(d). Upwards and downwards deviation from $m = 1$ indicate that $\Re$ is accelerated upon positive voltages and moderated upon negative ones with respect to $\Re_{\text{ini}}$, revealing a direct potential-induced electrostatic modulation of the reaction kinetics. We hypothesize that the exponent $m$ contains key information about the origin of this effect (e.g. an increase/decrease in surface coverage of negatively charged oxygen adsorbates), but further research is required to elucidate this point, *e.g.* by combining BCR with near ambient pressure XPS (NAP-XAS) studies.[61] Complementary, the close relation found between $\Re \propto j_{\text{sat}}$ and $\Re_{\text{ini}} \propto \Delta\sigma\tau^{-1}$ demonstrates that the two individual approaches for the mechanistic interpretation of the ORR, separately developed by Merkle[30] and Schmid,[31] are actually two sides of the same coin.

The driving force for net oxygen flux across the WE surface is a difference of the oxygen chemical potential of the WE and the atmosphere, $\Delta\mu_O = \mu_{\text{WE}} - \mu_{\text{atmo}}$. Within a first order kinetic regime the flux is directly proportional to $\Delta\mu \propto \Delta c = c_\infty - c(t)$. In classic chemical (relaxation) experiments, $\Delta c$ arises due to a change in the $pO_2$ (and thus $c_\infty$), while for polarization measurements the atmosphere is maintained ($c_\infty = $ const.) but the oxygen concentration of the WE, $c(t)$, is altered (given a surface limited overall reaction). Analogue to the definition of the chemical surface exchange coefficient, $k^\delta$, in Eq. 8, we can therefore express an electrochemical surface exchange coefficient, $k^{\text{p}}$, for polarization experiments. With the flux given by $J = \Re = (2e)^{-1}j_{\text{sat}}$ and the affinity, here $\Delta c$, taken from titration measurements via $\Delta c = \Delta\delta V_{\text{uc}}^{-1}$, we can write:

$$k^{\text{p}} = J \cdot \Delta c^{-1} = \frac{j_{\text{sat}} d_{\text{film}}}{\int (j(t) - j_{\text{sat}}) dt}. \qquad \text{Eq. 12}$$

This enables a direct comparison of chemical and electrochemical surface exchange coefficients from a single BCR experiment, as shown in Figure 4(e). Notably, for small absolute voltages (symbols with low opacity), we find almost matching $k$ values, *i.e.* points falling on to the diagonal. Increasing the bias magnitude affects mainly $k^{\text{p}}$, as marked with black arrows. This dependence becomes readily visible when plotting $k^\delta$ and $k^p$ against the applied voltage, as depicted in Figure 4(f) for 455°C and in SI-Figure 11 for all analysed temperatures. While $k^\delta$ is approximately constant, $k^p$ grows exponentially with $U_{DC}$, being smaller than $k^\delta$ under cathodic polarizations and significantly accelerated under anodic ones. Note that the crossing of the trend lines is slightly offset to below 0 V, which is expected to be an artefact due to setup related, small thermovoltages. As for the reaction rate above, we speculate that the drastic influence of the polarity on $k^p$ is linked to electrostatic effects, such as coverage changes of charged adsorbates and modifications of the surface dipole layer with $\Delta\chi(\eta) \neq 0$. Thus, these measurements provide direct evidence that a charge transfer reaction must take place before or during the RDS, as otherwise $\Delta\chi(\eta)$ should vanish (for the forward rate of the RDS) and $k^\delta = k^p$, independent of $\eta$ (within a limited $\eta$ range, see next section).

The temperature dependence of $k^p$, accessible by conventional $j-\eta$/titration measurements, is shown in SI-Figure 12, together with the $k^\delta$ values from relaxation. The corresponding activation energies, $E_A$, for different biases are added to Figure 3(c) (black diamonds). Particularly for oxidation steps (cathodic biases), activation energies obtained from relaxation transients ($k^\delta$) and titration experiments ($k^p$) match very well, pointing towards the same RDS. This demonstrates that electrochemical surface exchange coefficients and corresponding $E_A$ values can be retrieved from standard $j-\eta$ measurements.

Up to this point we can derive some general conclusions. The combination of electrical and chemical driving forces proved a powerful, yet simple approach to study all relevant materials descriptors for MIET reactions. BCR provides access to the bulk oxygen stoichiometry and electrical conductivities, as well as surface kinetic properties, such as net and initial reaction rates and the chemical and electrical surface exchange coefficients. Additionally, we showed that an electrochemical surface exchange coefficient of the WE and corresponding activation energies can be determined from traditional $j-\eta$ measurements, which also cover aspects of electrochemical impedance spectroscopy by containing kinetic information of the counter electrode (via the onset current density). Moreover, we verified the potential of the BCR technique to investigate electrostatic effects arising under current and to separate its contribution to the total reaction rate, opening up new opportunities for detailed studies of surface reactions, while highlighting that $\Delta\chi$ can play a significant role in $j-\eta$ measurements.

Regarding the investigated LSF|LSGM|Ag system, we can conclude that: (i) a first order WE surface reaction is limiting the overall electrochemical cell performance. (ii) Initial electronic and ionic bulk defect states ($\sigma$ and $\delta$) have only a minor effect on the exchange kinetics, in contrast to $pO_{2,ref}$, which strongly alters the activity. (iii) Already rather small positive/negative biases accelerate/deaccelerate the reaction rates, implying that net electric charge is transferred before or during the RDS. Based on these points, we hypothesize that charged oxygen adsorbates play a direct and dominant role in the RDS of the ORR/OER, while a more detailed mechanistic interpretation will be the subject of a future work.

### 3.3. Limit of the kinetic regime and experimental restrictions

To conclude the introduction of this novel technique, we look at physicochemical and experimental limitations of bias-triggered conductivity relaxation. The above analysed bias range was limited to ±100 mV, corresponding to changes of more than 3 orders of magnitude in effective $pO_2$. All resulting transients obeyed single exponential behaviour, a clear sign for a linear reaction order. Exceeding this bias range and thus triggering much larger $pO_2$ steps, we can analyse the validity range of the first order kinetic regime. In Figure 5(a) we analyse the relaxation from different cathodic polarization steps down to -800 mV. Down to -200 mV, transients can be well described using an exponential function (*i.e.* Eq. 9, see dashed lines in Figure 5(a)). Notably, an overpotential of -200 mV at this temperature corresponds to a change in $pO_2$ over six orders of magnitude, *i.e.*, from $10^{-3}$ mbar to $10^3$ mbar. In contrast to classical ECR experiments, BCR is not affected by any reactor flushtime and therefore even such large steps can be described within a linear kinetic regime, as indicated by the simple exponential behaviour.

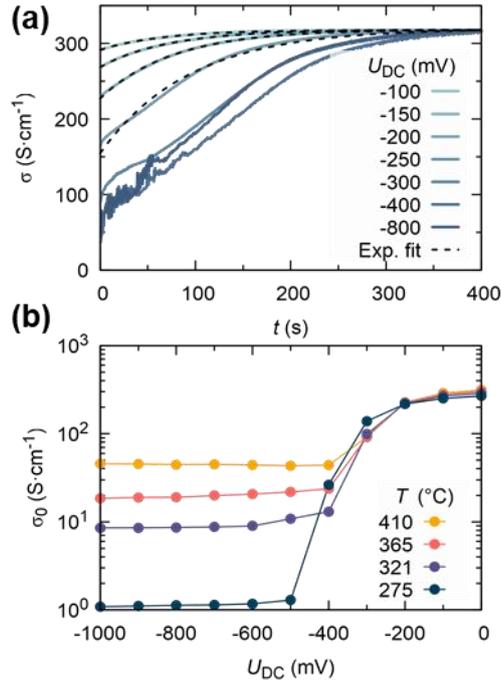

Figure 5: (a) LSF relaxation processes at 410 °C for large chemical driving forces: with increasing $U_{DC}$, transients deviate from the expected behavior for a linear kinetic regime of the oxygen surface exchange, *i.e.* following a simple exponential curve. (b) Temperature dependent saturation of the initial conductivity, $\sigma_0$, at strong cathodic polarizations.

Further decreasing the overpotential, deviations become gradually more severe and more complex - a clear sign of a collapse of the first order approximation. At -400 mV and below, the initial stage is highly turbulent and conductivity measurements become subject to noise, which was not observed previously. Where fitting was not applicable, the initial conductivity, $\sigma_0$, was determined from the first measured point after switching the electrode configuration. $\sigma_0$ saturates below a temperature dependent polarization level, as clearly seen in Figure 5(b). Notably, this saturation is independent of the atmospheric pressure, see SI-Figure 3(c). While transitions from predominantly hole conduction under oxidizing conditions to electron governed electrical conductivity under highly reducing conditions are commonly found for LSF and other members of the family of oxides, the here observed levelling off is rather ascribed to an experimental limitation. With increasing deviation from equilibrium, reaction rates can alter and other, previously neglectable kinetic contributions may become crucial, such as the reaction kinetics of the counter electrode, as schematically illustrated in SI-Figure 13. This reduces the efficiency of translating the applied voltage into an overpotential inside the WE. At this point, an increase of the voltage magnitude does not further change the oxygen stoichiometry of the MIEC, causing the observed saturation in conductivity. It is noteworthy that this experimental limitation for large polarizations occurs well beyond the linear kinetic regime and therefore does not impact the operational window of interest for BCR measurements.

## 4. Conclusions

We have developed bias-triggered conductivity relaxation (BCR) as novel tool to study kinetic processes and simultaneously retrieve thermodynamic insight into mixed ionic electronic conducting oxides. The combination of electrochemical titration and in-plane resistivity measurements and the replacement of steps in atmospheric $pO_2$ with equivalent electrochemical ones, was demonstrated to provide highly efficient access to all relevant parameters for the analysis of oxygen transport reactions and extend measurement flexibility and precision. We used BCR to perform a standard kinetic analysis of LSF thin films. A comparison of our results with existing literature was conducted to ascertain the validity of the technique. The unprecedented correlation of electrochemical and chemical measurements revealed the direct biased induced influence on reaction kinetics. Moreover, we uncovered that the electrochemical surface exchange coefficient, as well as a kinetic descriptor of the counter electrode, are hidden in conventional current density-overpotential measurements. We anticipate that the simple implementation and rich information depth of BCR, combined with experimental advantages over conventional techniques will result in rapid and widespread dissemination and foster the development of materials for sustainable energy applications.

## 5. Experimental section

$La_{0.6}Sr_{0.4}FeO_{3-\delta}$ (LSF) thin films of 120 nm thickness were grown at TU Wien by pulsed laser deposition (PLD) on 5×5 mm $La_{0.95}Sr_{0.05}Ga_{0.95}Mg_{0.05}O_{3-\delta}$ (LSGM) single crystal substrates (100).[62,63] Metallic Ti(5 nm)/Au (100 nm) top electrodes were fabricated in Nanofab clean room facilities using photolithography and evaporation. The width and spacing of the metallic stripes is $\approx 0.7$ mm. The porous Ag counter electrode was fabricated by brush-paint.

Phase and orientation of LSF thin films were characterised by X-ray diffraction (XRD) in the $\theta - 2\theta$ configuration using a Bruker D8 Advance series II diffractometer (CuKα radiation). The surface homogeneity was studied using an FEG ZEISS GeminiSEM 300 microscope in secondary and backscattered electron mode with an accelerating voltage of 3 kV in high vacuum. Image analysis was performed using ImageJ and the Trainable Weka Segmentation plugin.

Functional characterisation was performed using current-overpotential, four-point electrical resistivity and electrochemical impedance spectroscopy (EIS) measurements using a Keithley 2400 sourcemeter and a Solartron 1260 impedance analyser. For *in situ* measurements, the sample was placed onto a 1/2" ceramic heating stage of a high temperature cell (Nextron), equipped with six electrical probes. A Si substrate with an Ag paint-brush layer was placed in between heater and sample to provide electrical contact to the counter electrode. The temperature at sample position was calibrated beforehand using a Pt100 thermocouple. However, a certain uncertainty in sample temperature is inherent to such type of setup, as well as a temperature gradient across the cell. Annealings were performed in a dynamic, dry, mixed $O_2/N_2$ atmosphere at 1 atm and constant flow rate of 200 ml/min, whereas the oxygen partial pressure was controlled via the flow ratio of pure oxygen to nitrogen gas. The $pO_2$ was recorded at the exit of the chamber using a Rapidox 2100 gas analyser. The full setup (including cell temperature, atmosphere, electrical sample configuration via a digital relay and different measurement types, incl. polarization, conductivity, EIS) was automatically controlled and monitored using a home written Labview program.

## Acknowledgments

We acknowledge M. Burriel and C. Jiménez (both LMGP) for continuous scientific exchange and access to experimental setups. A.St. is grateful for valuable discussions with R. Merkle (Max Planck Institute for Solid State Research). This research was funded in part by the Austrian Science Fund (FWF) 10.55776/PAT2412325. Authors acknowledge NANOFAB facilities of Institut Néel. This research has benefited from characterization equipment of the Grenoble INP - CMTC platform supported by the Centre of Excellence of Multifunctional Architected Materials "CEMAM" n°ANR-10-LABX-44- 01 funded by the "Investments for the Future" Program. Last but not least A.St. would also like to express his deep gratitude to A. Richard (Ville de Grenoble) for the magic you have brought into my life.

## Data Availability Statement

The data that support the findings of this study will be made available upon first request at 10.5281/zenodo.18335600.

## Declaration of Competing Interest

The authors declare no known conflict of interest.

## Author Contributions

A.St. developed the original concept and methodology, built the experimental setup, performed the measurements, analysed the data and wrote the manuscript. A.Sch. and A.St. prepared the samples. A.R. assisted the measurements. A.St, A.Sch. and J.F. critically discussed the results. A.B., A.S and J.F. acquired the funding for research and personnel. All co-authors contributed to the revision of the manuscript.

# Supplementary Information:

**Supplementary Information Note 1: Sample and setup requirements for BCR**

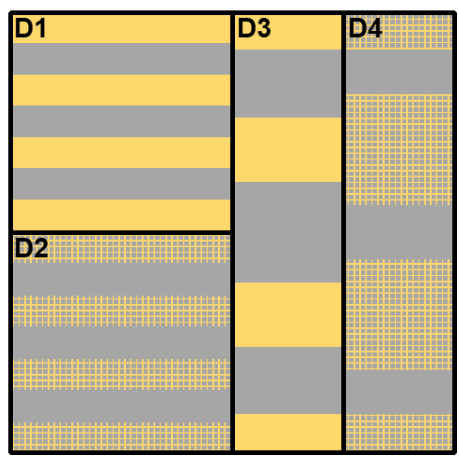

SI-Figure 1: 10×10 mm sample with four different device geometries with open grid (D2, D4) and closed layer (D1, D3) metallic top electrodes for testing purposes. In this work, squared, 5×5 mm devices of type D1 with closed layer electrodes were used.

Obtaining meaning full data via the BCR technique based on the combination of polarization and conductivity measurements requires specific materials properties and precautions in the electrode design.

Homogenous polarization of the working electrode and effective current collection is limited to an area in proximity to the metallic top electrodes. This area depends on the ratio of in-plane electrical conductivity and out-of-plane oxygen activity of the WE, defining a material-specific screening length. On the other hand, four-point electrical resistivity measurements require a certain electrode spacing to provide adequate results. Furthermore, oxygen exchange is blocked underneath the metallic current collector. Thus, switching between open grid and closed layer designs and different sample geometries, as shown in SI-Figure 1, allows to vary the accessible WE surface and adapt for its total exchange activity. This gives a parameter to tune the ratio of open surface WE activity and full area CE activity and ensure WE surface reactions are rate limiting (as long as MIEC ionic bulk diffusivity and electronic conductivity are fast enough to compensate for longer in-plane diffusion lengths).

Sufficiently high in-plane electronic conductivity of the MIEC is not only important for its homogeneous polarization during step 1, but also for precise conductivity measurements during step 2: a high ratio of electronic (MIEC) vs. ionic (MIEC and electrolyte) in-plane conductivity avoids parallel ionic current pathways through the electrolyte (or the MIEC itself), which would interfere with the accurate modelling of the relaxation transients for the determination of kinetic parameters.

Note, that the presence of the current collector may influence the observed kinetics, based on catalytic influences, contaminant interactions, introduction of impurities, space charge modifications and surface blocking effects. However, these effects are expected to be equally relevant during polarization as well as relaxation processes and thus, are not thought to contribute to the observed differences.

Temperature gradients across an electrochemical cell, as for example common for environmental temperature cells with asymmetric heating, can produce thermovoltages in the range of tens of mV. This not only affects small polarization steps, but also leads to unequal equilibrium states for the un-biased but closed-circuit out-of-plane polarization configuration and the in-plane electrical resistivity configuration and has to be considered in the calculation of $pO_{2,\text{eff}}$.

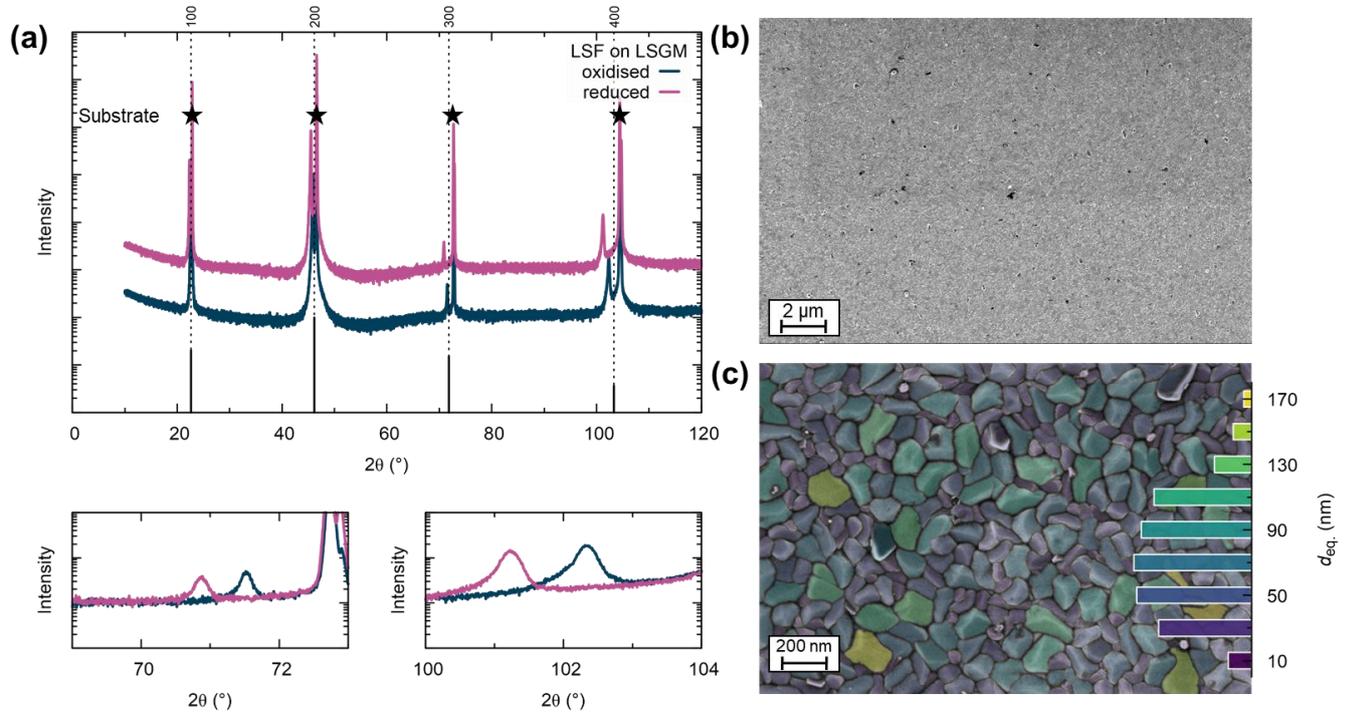

SI-Figure 2: (a) X-ray diffractograms of fully oxidised and reduced $(La,Sr)FeO_{3-\delta}$ thin films grown on (100) $La_{0.95}Sr_{0.05}Ga_{0.95}Mg_{0.05}O_{3-\delta}$ (LSGM) single crystal substrates. The LSF films exhibit a pseudo-cubic structure with (h00) orientation (ICDD: 00-069-0127). The lattice parameter is sensitive to the oxygen stoichiometry, as clearly observable in the magnified regions in the two lower panels for the fully reduced and fully oxidised films, achieved by annealing in low and high $pO_2$, respectively. (b) Low and (c) high magnification scanning electron microscopy images (secondary electron mode) of LSF top surface. The false colour in (c) indicates the grain size, with the corresponding histogram of the distribution of the equivalent diameter (assuming circular grains), with an average $d_{eq} \approx 80$ nm.

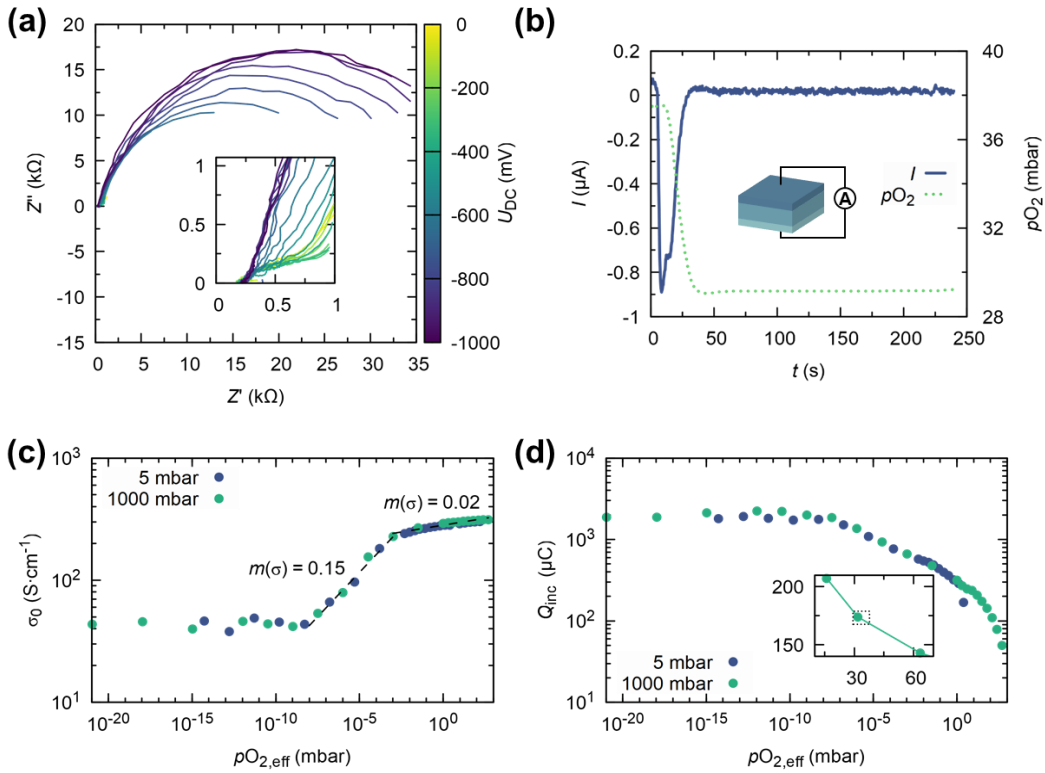

SI-Figure 3: (a) Electrochemical impedance spectroscopy (EIS) 410 °C in 1 bar of $O_2$. (b) A gas-sensor type test, where current is measured across an electrochemical device upon a change in $pO_2$. (c) The initial conductivity, $\sigma_0$, and (d) incorporated charge, $Q_{\text{inc}}$, obtained via bias-triggered conductivity relaxation (BCR) measurements, as function of the effective oxygen partial pressure. Note that $Q_{\text{inc}}$ is not an absolute measure. Therefore, the data set for 5 mbar was shifted upwards to account for the different starting point in oxygen off-stoichiometry.

**Supplementary Information Note 2: verification of WE surface limited regime**

SI-Figure 3 presents the experimental verification that a working electrode (WE) reaction is overall rate limiting and consequently an applied bias is (almost) fully translated into an electrochemical overpotential using four different approaches. Electrochemical impedance spectroscopy (EIS) data for a LSF thin film, obtained at 410 °C in 1 bar of $O_2$, is shown in SI-Figure 3(a). The very large semi-circle corresponds to the LSF electrode. The inset reveals the very minor electrolyte and counter electrode (CE) polarization contributions of approximately 200 Ω (< 1 %) and 500 Ω (≈ 1-2 %), respectively.

The initial conductivity, $\sigma_0$, and incorporated charge, $Q_{\text{inc}}$, obtained via bias-triggered conductivity relaxation (BCR) measurements are shown in (c) and (d), respectively, as function of the effective oxygen partial pressure. The preservation of the same curvature and overlap of data points measured at different oxygen pressures inside the chamber ($pO_{2,\text{ref}}$), confirm that the applied bias is fully converted into an overpotential, which modifies the oxygen chemical potential of the WE. The levelling off of $\sigma_0$ and $Q_{\text{inc}}$ at low $pO_{2,\text{eff}}$ ($\lesssim 10^{-10}$ mbar) is likely a measurement artefact, as discussed in the main text.

Additionally, we introduce a gas-sensor type test, as shown in SI-Figure 3(b), for the convenient confirmation of WE surface limitation. Here, WE and CE are electrically short-circuited using an ampere meter and the out-of-plane current is measured upon performing a switch in the gas atmosphere. The negative current indicates that the mixed ionic electronic conducting working electrode equilibrates (at least partially) to the new, lower $pO_2$ through the counter electrode, *i.e.* the oxygen evolution reaction is outsourced to the CE. Some oxygen may still be incorporated through the native WE surface.

But that this amount is small can be verified by comparing the area under the $I(t)$ curve with $Q_{\text{inc}} = -12\ \mu C$ with the dependence of $Q_{\text{inc}}$ con the effective $pO_2$, as shown in SI-Figure 3(d). The dotted rectangle in the inset of (d) marks the step as performed in (b). By linear interpolation we obtain an expected change of $Q_{\text{inc}} = -11\ \mu C$ (height of the rectangular). Within the error of this analysis, this figure is ident to the value reported above, confirming that most of the oxygen required for the stoichiometric change of the WE is incorporated through the CE. Note that as the $pO_2$ sensor is mounted at the gas exit of the sample chamber, a delay in the signal can be observed in (b). On the other hand, the finite transition width, due to the low oxygen gas flow used here (200 ml min$^{-1}$), would indicate a potential flush time limitation for conventional ECR measurements for samples with fast kinetics.

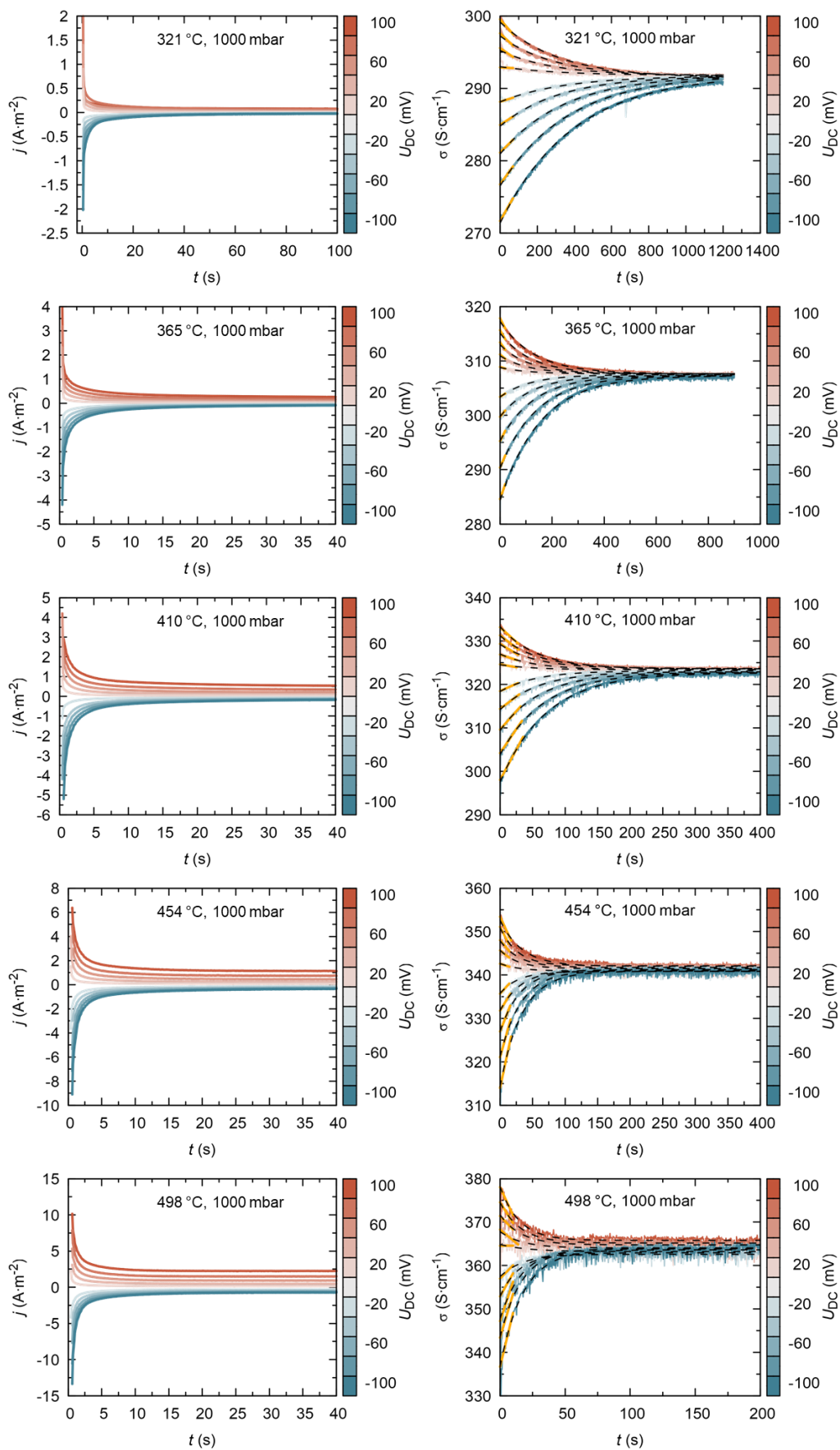

SI-Figure 4: Bias-triggered conductivity (BCR) measurements: polarization (left) and subsequent relaxation (right) curves at different temperatures and various cathodic and anodic biases.

**Supplementary Information Note 3: Counter electrode + electrolyte kinetics**

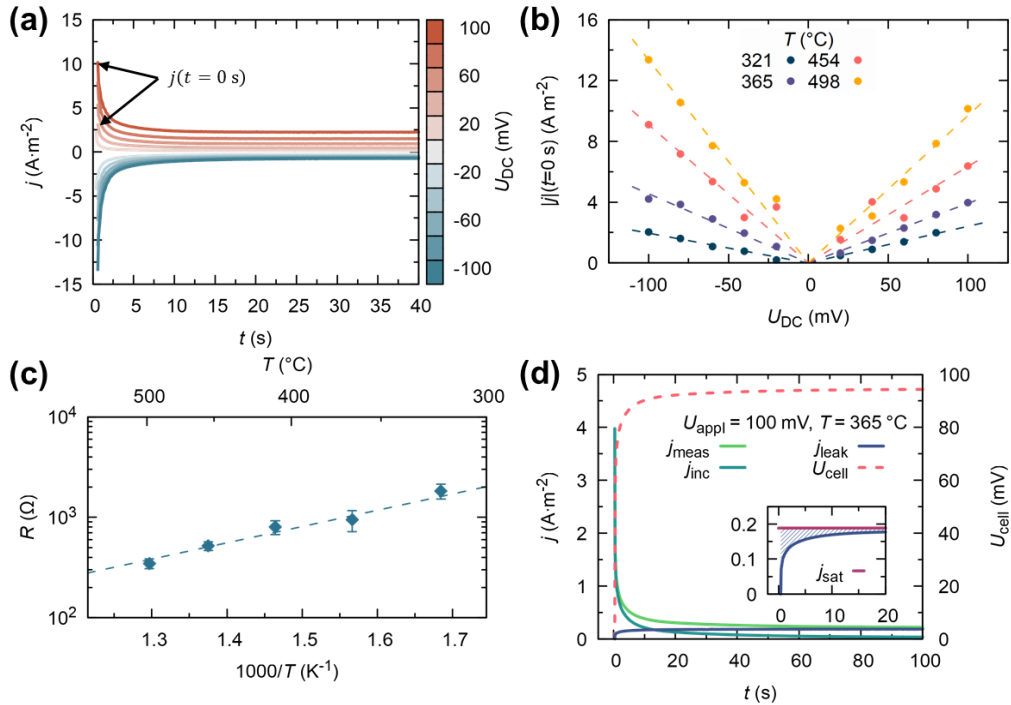

SI-Figure 5: (a) $j(t)$-curves with marked onset current density, $j(t = 0 \text{ s})$. (b) Ohmic characteristic of $j(t = 0 \text{ s})$ at various temperatures. (c) Arrhenius plot of entangled electrode and electrolyte resistance, obtained from (b). (d) Analysis of time evolution of cell voltage (right $y$-axis) and different current contributions (left $y$-axis). The marked area in the inset highlights the *forgotten* charge, when assuming a constant leakage current for all $t$.

$j(t)$-curves, as shown in SI-Figure 5(a), are commonly evaluated in terms of saturation current densities as well as transported charge (*i.e.* the area between $j_{\text{meas}}(t) - j_{\text{sat}}$). Here we use additionally the onset current density, $j(t = 0 \text{ s})$, as a measure of the electrochemical activity of the counter electrode and the electrolyte and further, to evaluate the two different current contributions: namely the one leading to a change in the oxygen stoichiometry ($j_{\text{inc}}(t)$) and the current density corresponding to the leakage flux of oxygen through the WE surface. The $j(t = 0) - U$ plot shown in SI-Figure 5(b) reveals an ohmic characteristic. Thus, the slope is inversely proportional to a resistance. At 450 °C, this resistance matches very well the sum of electrolyte and counter electrode contributions as obtained by EIS, see SI-Figure 3(a), and therefore supports our interpretation of $j(t = 0 \text{ s})$. The temperature evolution of $R$ is shown in SI-Figure 5(c).

Using the linear $j(t = 0) - U$ dependence (with the slope $m$) from SI-Figure 5(b) and the Butler-Volmer-type dependence of the saturation current density, as presented in the main text, we can approximately analyse the process of building-up the cell voltage, *i.e.* the overpotential in the WE. The overpotential equals the applied voltage minus a voltage drop due to the flow of a current through the counter electrode and the electrolyte, thus we can write: $\eta(t) = U_{\text{cell}}(t) \approx U_{\text{appl}} - U(j(t))$, with $U(j(t)) = j(t)/m$ obtained from the dependence shown in (b). The time evolution of the overpotential (pink dashed line) is shown using the right $y$-axis in SI-Figure 5(d). It sharply rises at $t = 0$ s and saturates at around 95 mV (*i.e.* 95 % of the applied voltage). This simple method allows to quantify the actual overpotential in the presented sample, without the need for additional characterisation steps, such as electrochemical impedance spectroscopy (EIS).

The rising WE overpotential corresponds to a step in oxygen chemical potential across the WE surface, which creates a driving force for net oxygen flux and therefore triggers a leakage current, $j_{\text{leak}}(t)$. The magnitude of the leakage current can be estimated based on the actual cell voltage, $U_{\text{cell}}(t)$, using the Butler-Volmer equation. The time evolution of the leakage current density is shown in (d) and magnified in the inset. Compared to assuming a constant leakage current density (i.e. $j_{\text{leak}}(t) = j_{\text{sat}}$), a more precise incorporation current density can be defined via $j_{\text{inc}}(t) = j_{\text{meas}}(t) - j_{\text{leak}}(t)$. Thus, the marked area in the inset of SI-Figure 5(d), between $j_{\text{leak}}$ and $j_{\text{sat}}$, corresponds to additional charge incorporated into the MIEC, which was previously not considered. For the LSF system studied here, the correction was found to be in the range of a few %.

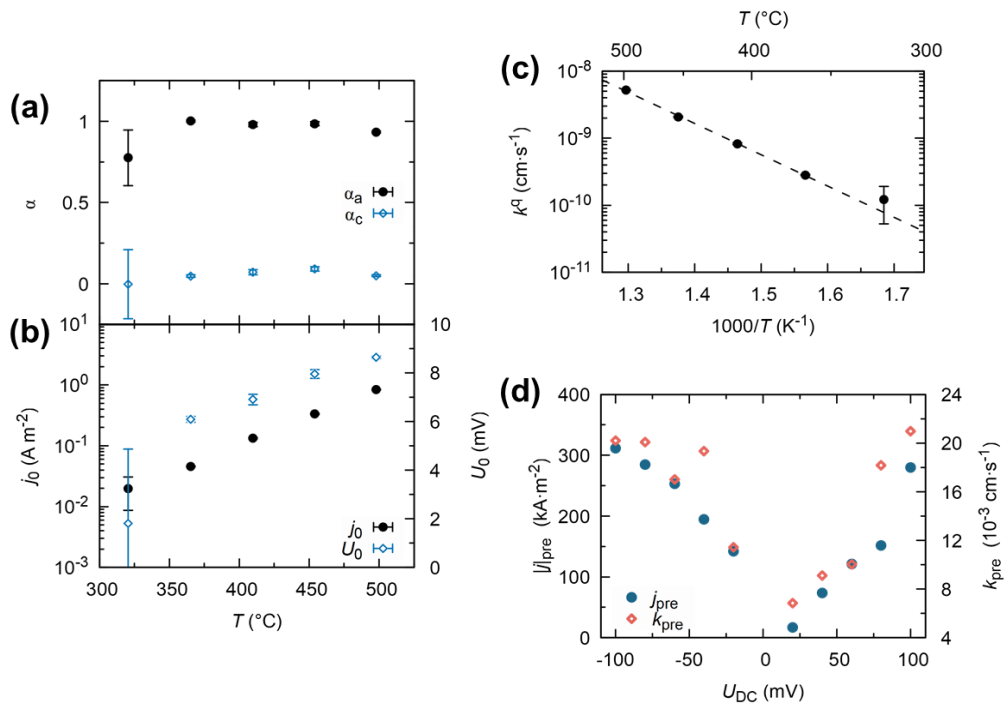

SI-Figure 6: (a & b) Fit parameters for the exponential Butler-Volmer type equation at different temperatures. (c) Arrhenius plot of electrical surface exchange coefficient, $k^q$. (d) Bias dependence of the pre-exponential factors from Arrhenius-law fittings of the saturation current (left y-axis) and the surface exchange coefficient (right y-axis).

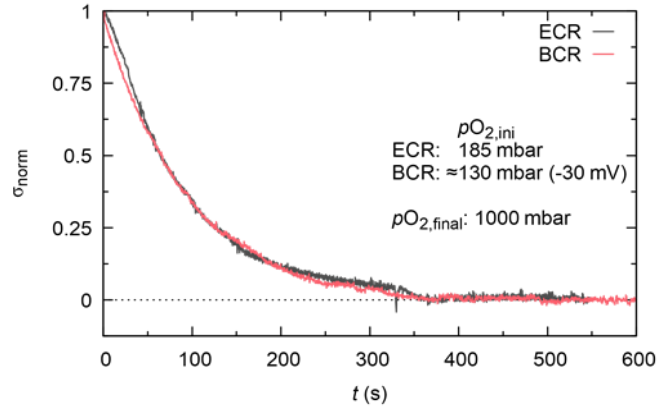

SI-Figure 7: Normalized conductivity relaxation transients obtained from standard ECR measurement via a change in atmospheric $pO_2$ and via novel bias-triggered conductivity relaxation, whereas the initial $pO_{2,\text{eff}}$ inside the MIEC was set via an applied voltage across the electrochemical cell at $t < 0$.

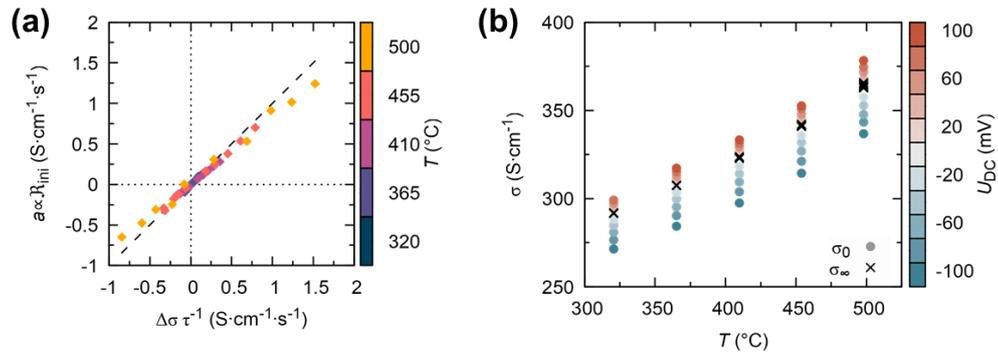

SI-Figure 8: (a) Correlation between the linear slope, $a$, of the initial stage of the relaxation transients and the ratio of the change in electrical conductivity, $\Delta\sigma$, and the exponential saturation time, corresponding to the linear term of the series expansion of the exponential fitting curve. The dashed line marks identity. (b) Temperature dependence of the electrical in-plane conductivity under different polarizations in 1 bar of oxygen.

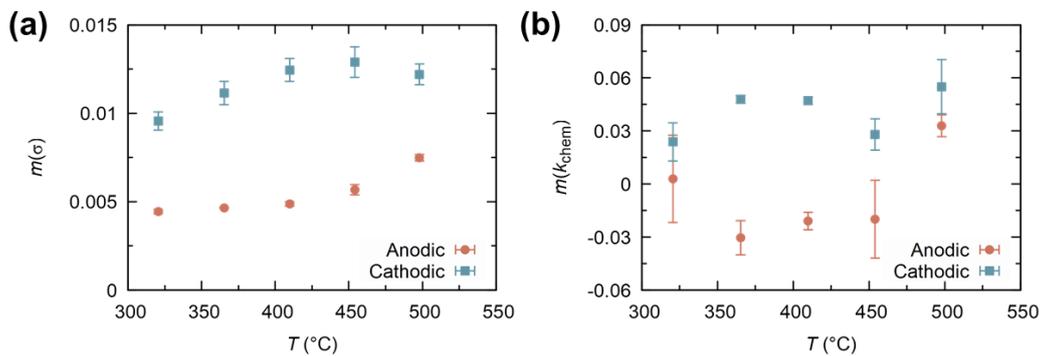

SI-Figure 9: Pressure dependence of (a) the electrical conductivity and (b) the surface exchange coefficient of LSF thin films at different temperatures for anodic and cathodic polarizations.

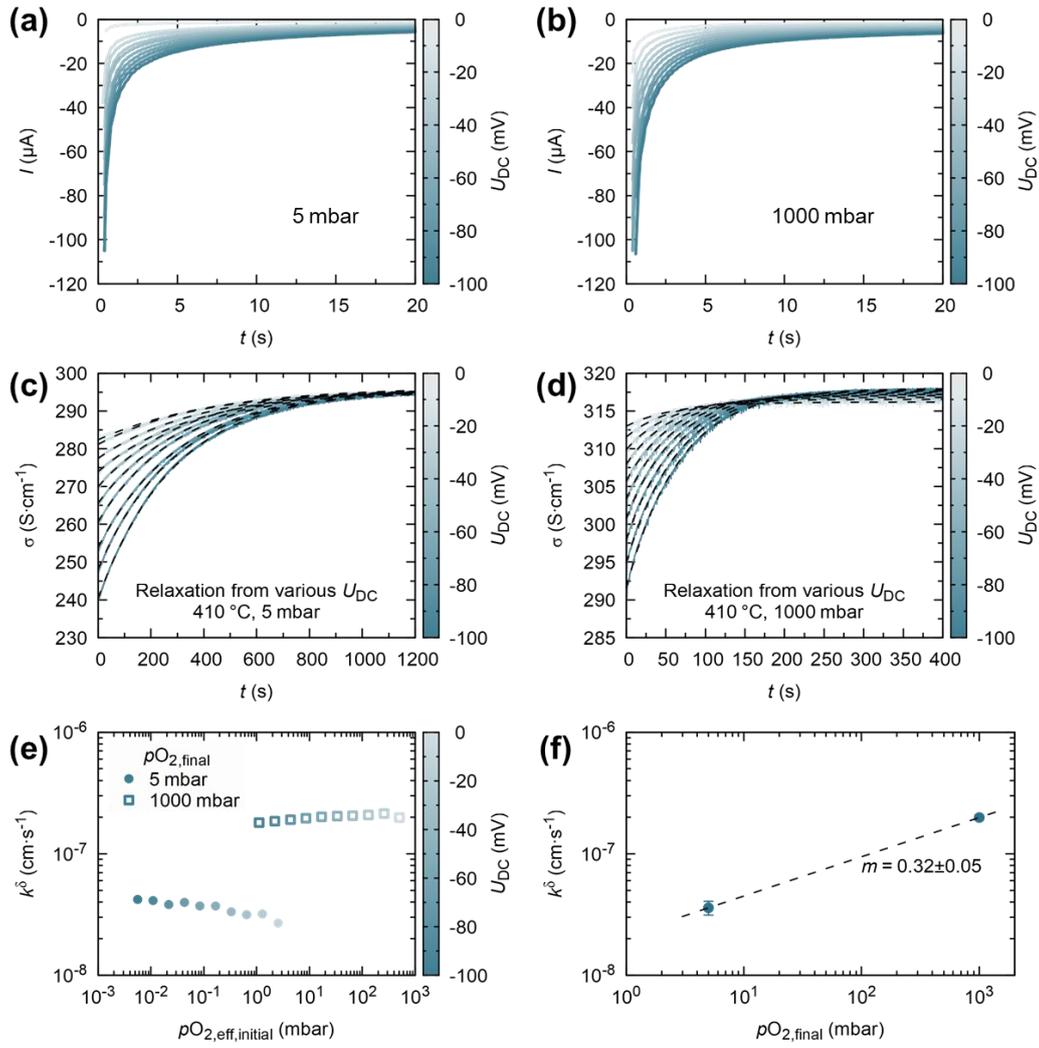

SI-Figure 10: (a & b) Polarization and (c & d) subsequent relaxation steps at 5 and 1000 mbar, respectively at 410 °C. (e) The chemical surface exchange coefficient, extracted from (c & d), is almost invariant to the step size but it is strongly lowered with decreasing $pO_{2,\text{final}} = pO_{2,\text{atmosphere}}$ inside the chamber. The $pO_2$ reaction order of the chemical surface exchange coefficient is approximately 0.3 as shown in (f).

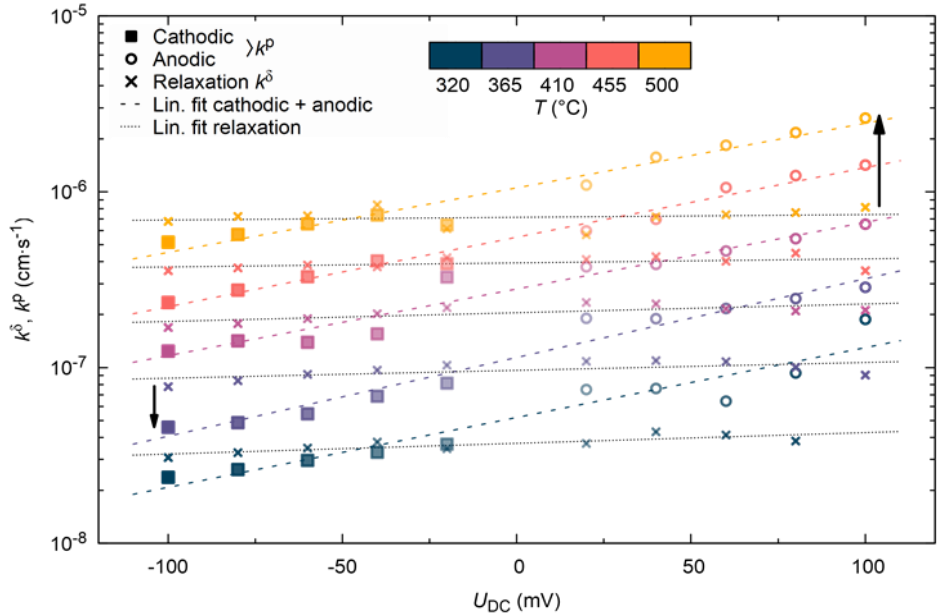

SI-Figure 11: Bias dependence of the electrochemical and chemical surface exchange coefficients, $k^p$ and $k^\delta$, obtained from $j-\eta$ measurements and the subsequent relaxation processes. Under cathodic polarizations, $k^p$ is slower compared to $k^\delta$, while positive voltages accelerate the surface exchange coefficient compared to the bias-free relaxation process, as marked with black arrows. $k^p$ varies exponentially with voltage (dashed lines), while $k^\delta$ is independent of the preceding voltage (dotted lines). The crossing of the two linear trends are observed for all temperatures below 0 V, which is likely linked to thermovoltages due to the deployed temperature cell and asymmetric heating.

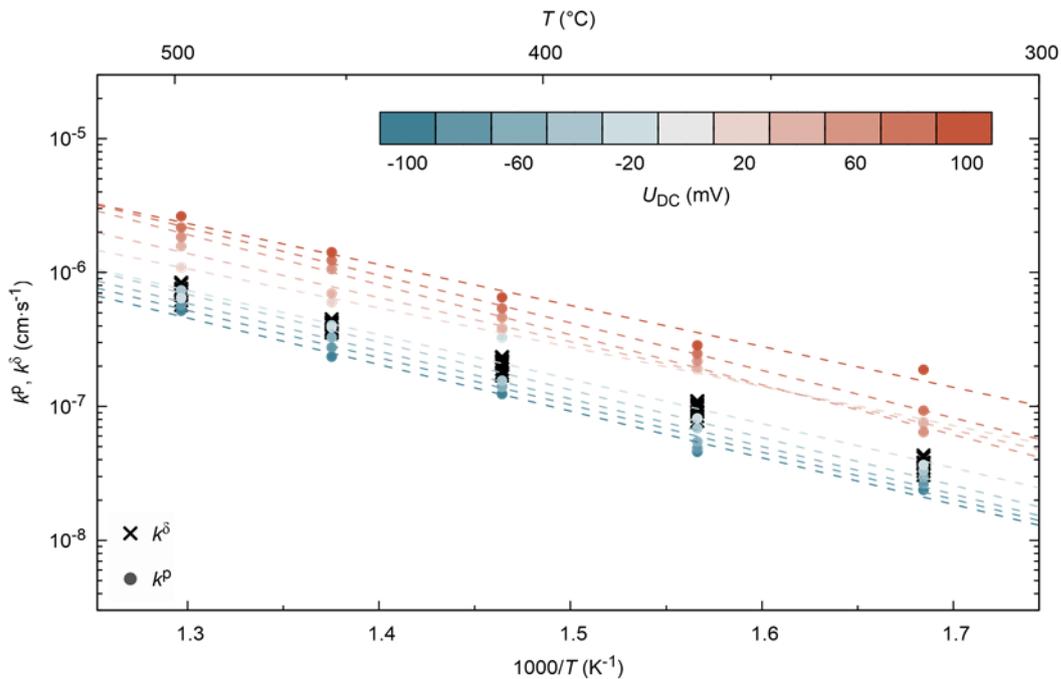

SI-Figure 12: Arrhenius plot of the electrochemical surface exchange coefficient, $k^p$, obtained via titration and $j-\eta$ measurements for different polarizations, compared to the chemical surface exchange coefficient, $k^\delta$ from relaxation processes. The activation energies are compared in the main text as function of the polarization.

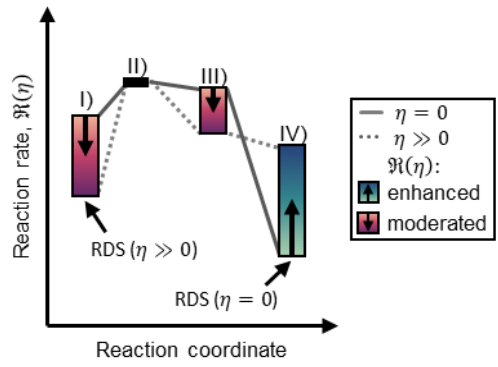

SI-Figure 13: Evolution of reaction rates under applied bias for an exemplary electrochemical cell with (I) oxygen reactions at the counter electrode, oxygen ionic diffusion through (II) the electrolyte and (III) the MIEC thin film and (IV) MIEC surface reactions. In the illustrated example, the RDS changes from step (IV) at 0 bias to step (I) under strong anodic polarizations.